\documentclass[journal, onecolumn]{IEEEtran}

\usepackage{cite}
\usepackage{amsmath,amssymb,amsfonts}
\usepackage{textcomp}
\usepackage{xcolor}
\usepackage{comment}

\usepackage{algorithm}
\usepackage{algorithmic}
\usepackage{microtype}
\usepackage{graphicx}
\usepackage{subfigure}

\usepackage{wrapfig}

\usepackage{mathtools}

\usepackage{cleveref}
\usepackage{url}

\begin{document}

\title{Performance and Energy Consumption of Parallel Machine Learning Algorithms}

\author{Xidong Wu,
        Preston Brazzle,
        and Stephen Cahoon%
}

\markboth{ECE 2166, February~2023}%
{Shell \MakeLowercase{\textit{et al.}}: Bare Demo of IEEEtran.cls for IEEE Journals}

\maketitle

\section{Abstract}

Machine learning models have achieved remarkable success in various real-world applications such as data science, computer vision, and natural language processing. However, model training in machine learning requires large-scale data sets and multiple iterations before it can work properly. Parallelization of training algorithms is a common strategy to speed up the process of training. However, many studies on model training and inference focus only on aspects of performance. Power consumption is also an important metric for any type of computation, especially high-performance applications. Machine learning algorithms that can be used on low-power platforms such as sensors and mobile devices have been researched, but less power optimization is done for algorithms designed for high-performance computing. 

In this paper, we present a C++ implementation of logistic regression and the genetic algorithm, and a Python implementation of neural networks with stochastic gradient descent (SGD) algorithm on classification tasks. We will show the impact that the complexity of the model and the size of the training data have on the parallel efficiency of the algorithm in terms of both power and performance. We also tested these implementations using shard-memory parallelism, distributed memory parallelism, and GPU acceleration to speed up machine learning model training.

\section{Introduction}
Machine learning is a class of data-driven algorithms and models where models progressively improve as they gain experience. It has many applications from image classification to robot control \cite{machine_learning}. By providing a set of training data, models can train themselves to accurately process new data outside of the training set. However, as the scale of training data and models has increased, it has become necessary to develop parallel versions of these algorithms to handle these larger problem sizes. Supervised learning algorithms are a highly parallelizable class of machine learning algorithms.

Supervised machine learning algorithms find the optimal parameters for a model when processing a predefined set of training data. This optimization problem can be solved in many unique ways, but some algorithms have become more popular than others. Stochastic gradient descent (SGD) is one of the most popular optimizers used to train machine learning models \cite{Ketkar2017}. A variant of gradient descent, SGD allows models to train on randomly sampled subsets of the training data each iteration rather than processing the entire set every time, which is relatively fast, computationally cheap, and parallelizable. 
In addition, We will also consider the genetic algorithm to train models, since it is highly parallelizable. The genetic algorithm keeps track of a population of several models, scoring each one on effectiveness and creating new generations from the highest-ranking individuals. Since the genetic algorithm scores many models at once, it can be easily split into batches for multithreaded processing.

Even with these techniques, training machine learning models with large parameters can require vast amounts of computational power and training time. Parallel machine learning algorithms can overcome these limitations by speeding up computation \cite{UPADHYAYA2013284}. Both shared-memory parallelism and distributed parallelism can leverage parallel computing to speed up the machine learning training process \cite{async_sgd}. In addition, distributed parallelism can divide the training workload across multiple machines, each with its own memory and processing power, enabling the training of large-scale machine learning models on massive data sets. 

However, as the size of data sets and models increase, so too does the consumption of energy. Energy analysis for machine learning algorithms is an emerging research area, but it has been looked into as far back as 1996 \cite{garcia-martinEstimationEnergyConsumption2019}. While much of the current work in machine learning focuses on the performance improvements gained, little attention is given to the energy consumed by different techniques and technologies. Work that compares these different implementations by both their speedup and energy consumption can be useful for designers and consumers alike. For example, while parallel training algorithms can have large speedups in comparison to their serial counterparts, the overhead costs for parallelism can add up as work is divided out between many processors. By finding the point at which power and performance efficiency diverges, researchers and consumers can have a better understanding of the trade-offs associated with improved performance.

In this paper, different machine-learning algorithms will be parallelized and trained with a variety of methods. Each of these methods will have both its performance and power compared to a serial baseline, and conclusions about these values will be drawn. Before the results are explained, the background of the topics in the paper and the methods used must be discussed.

\section{Background}

\subsection{Machine Learning Models}

\subsubsection{Artificial Neural Networks}

In an artificial neural network (ANN), data is fed into a series of nodes (or neurons) that calculate a linear combination on the data and send it to the proceeding layer. The coefficients of this combination are referred to as weights. Since nodes within a layer are not dependent on each other, their computation can be easily parallelized. The final layer represents the result of the model, and it often is a series of probabilities. For example, in an image classification model, each node in the output layer could represent the probability that the image belongs to a certain category. 

The training process involves finding the ideal sets of weights for the model to interpret data most effectively. Properly trained ANNs have been proven to be capable of universal function approximation \cite{ann_survey}. By providing an ANN model and a set of training data (inputs with expected outcomes), a machine learning algorithm can optimize the model to fit the training data. However, ANNs will often have difficulties extrapolating when given inputs outside of the training set. ANNs that perform well with training data but poorly with other data are considered over-fitted.

\begin{figure}[ht]
\vspace{-10pt}
\centering
\subfigure[Basic ANN]{
\hspace{0pt}
\includegraphics[width=.4\textwidth]{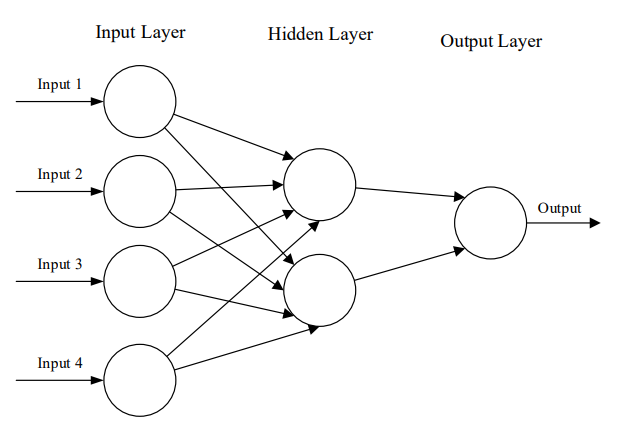}
}
\qquad
\subfigure[CNN]{
\hspace{0pt}
\includegraphics[width=.4\textwidth]{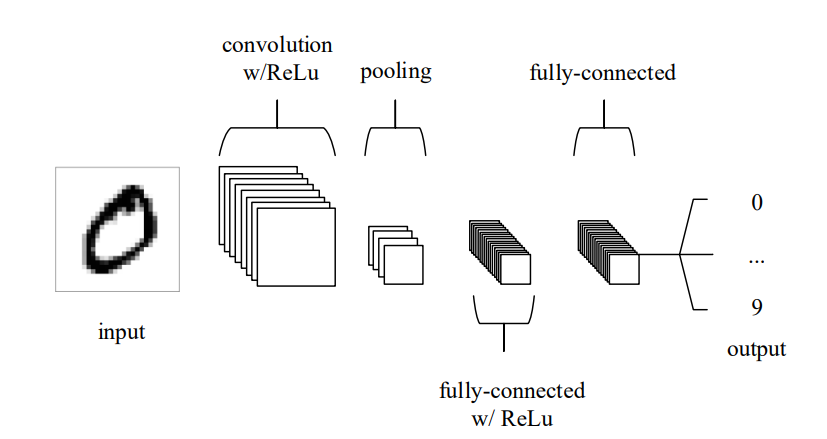}
}
\caption{A visualization of an ANN (a) and a CNN (b) from \cite{cnn_intro}. ReLu is shorthand for rectified linear unit.}
\vspace{-10pt}
\label{fig:stoch2}
\end{figure}

\subsubsection{Convolutional Neural Networks}

While ANNs are very useful, their computational complexity increases exponentially with larger inputs such as images. As ANN models grow in size, they also risk overfitting to the training data. Convolutional neural networks (CNNs) were proposed as the solution to these problems \cite{lecun2015deep}. CNNs are designed to process images effectively because  convolutional kernels could summarize the adjacent information and different layers could be regarded as multi-scale feature extractors.

There are three types of layers in a CNN: convolutional layers, pooling layers, and fully-connected layers. Convolutional layers perform several convolution operations on the input image to identify specific features. After training, the processed images are referred to as feature maps. Pooling layers simply downsample the result of the previous layer to reduce the computational load for proceeding layers. For example, a max pooling layer may replace each 2x2 grid of pixels with a single pixel containing the maximum value of the grid. This reduces the data size significantly, thus reducing computation requirements for subsequent layers. Fully-connected layers work similarly to ANNs. It will simply calculate a linear combination of all inputs and apply an activation function. A common activation function is ReLu, which replaces any negative values with 0. The fully-connected layer produces the overall output of the model \cite{cnn_intro}. \cite{he2016deep} about Deep Residual Networks (ResNets) has been one of the most cited paper (citation 162,170) in all research areas

\subsubsection{Logistic Regression}
Logistic regression is a classical machine learning model used in binary classification tasks. Logistic regression is a simple yet powerful algorithm that is widely used in various applications such as credit risk modeling, marketing analytics, and medical research. It is also often used as a baseline to compare against more complex models in research and development of new algorithms. It uses a logistic function to model the probability of a binary response variable. Logistic regression models are trained by finding the parameters that maximize the likelihood of the observed data. The output of this function represents the probability of the binary response variable, given the predictor variables. The logistic function is listed below: 
\begin{align}
p(x)=\frac{1}{1+e^{-\left(\beta_0+\beta_1 x\right)}}   
\end{align}
This resulting probability represents an 'is' or 'is not' relationship between the input data and the resulting labels. For example, if a logistic regression model were trained to label people with a certain marital status as 'happy' or 'unhappy', it would predict the probability that person is 'happy' given their marital status  \cite{demaris}. Thus, the model would classify the data to the category with the highest likelihood.

Logistic regression can be expanded to predict outcomes with many categories. This is often referred to as multinomial logistic regression \cite{multinom_log}. In multinomial logistic regression, a likelihood is calculated for every possible outcome category, and the model assigns the input data to the category with the highest outcome.

\subsection{Model Training Methods}

\subsubsection{(Stochastic) Gradient Descent}
Gradient descent can be used to find optimal parameters for a predefined loss function. In the case of supervised machine learning, the loss function is normally the difference between the model output and the expected output for the set of training data. By finding the gradient of the loss function, we can update the parameters in that direction to follows the path of steepest descent. The process of repeatedly calculating the gradient and traveling a small distance in that direction is known as gradient descent.

One of the main drawbacks of gradient descent is that the entire set of training data is used to find the gradient, which is computationally expensive and slow. Rather than using the full set, stochastic gradient descent randomly selects a subset of the training data to use at every step. Although this reduces the overall accuracy and adds noise into the gradient direction, stochastic gradient descent can progress at a much higher speed \cite{Ketkar2017}. SGD algorithm is summarized in algorithm \ref{alg:1}, where $f(\cdot)$ is the loss function and $\mathcal{D}$ is the training datasets.

\begin{algorithm}[tb]
\caption{Stochastic gradient descent (SGD) Algorithm}
\label{alg:1}
\begin{algorithmic}[1] 
\STATE {\bfseries Input:} Training epoch number $T$, learning rate $\eta$; and mini-batch size $b$; \\
\STATE {\bfseries Initialize:} Model parameters $x_{0}$;
\FOR{$t = 1, 2, \ldots, T$}
\STATE  Draw mini-batch samples $\mathcal{B}_{t}=\{\xi^j\}_{j=1}^{b}$ with $|\mathcal{B}_{t}|=b$ from training datasets $\mathcal{D}$\\
\STATE
$x_{t} = x_{t-1} - \eta \nabla_x f(x_{t-1}; \mathcal{B}^i_{t})$\\
\ENDFOR
\STATE {\bfseries Output:} $x_T$.
\end{algorithmic}
\end{algorithm}

As the size of models and datasets increase, there is a rise of distributed training in deep learning \cite{dean2012large}. In recent years, synchronous and asynchronous training have been successfully applied to many optimization algorithms such as stochastic gradient descent (SGD) \cite{recht2011hogwild, lian2015asynchronous} and SVRG \cite{huo2017asynchronous, bao2022doubly}. \cite{recht2011hogwild} parallelizes SGD and proves parallel computing can successfully speed up machine learning training. \cite{lian2015asynchronous} studied the asynchronous parallel implementations of SGD for nonconvex optimization (deep learning models) and provide theoretical support. \cite{huo2017asynchronous} utilized variance-reduced technology to accelerate the convergence of distributed training. Distributed optimization has obtained successes in various deep learning domains, such as computer vision \cite{goyal2017accurate}, generative modeling \cite{brock2018large} and reinforcement learning \cite{zhang2021taming}. Some communication-efficiency distributed methods are also considered \cite{wu2022faster, wu2023decentralized}. However, most of them focus on synchronous training and does not compare different  (synchronous and asynchronous, shared-memory and distributed) parallel ways to speed up model training.

The SGD algorithm can be parallelized as synchronous or asynchronous. In synchronous parallel SGD, we use parallel computing in each iteration. In shared-memory parallel computing, we use the multi-processing method to speed matrix multiplication up. We use parallel each iteration with "\#pragma omp parallel for" to update models.  In distributed parallel computing, each processing node samples data points locally and calculates its own stochastic gradient. A server then aggregates these gradients from each client and updates the model with the overall gradient. Finally, the server sends this updated model to each processing node. On the other hand, asynchronous training does not require a synchronization step and avoids model-update bottlenecks. In shared-memory parallel computing, we make each iteration parallel. In distributed parallel computing, when the server receives a stochastic gradient, it updates the model immediately and sends it back only to the processing node that calculated that gradient. 

\subsubsection{Genetic Machine Learning}

\begin{figure}[ht]
\vspace{-10pt}
\centering
\subfigure[how selected solutions are combined for future generations
]{
\hspace{0pt}
\includegraphics[width=.5\textwidth]{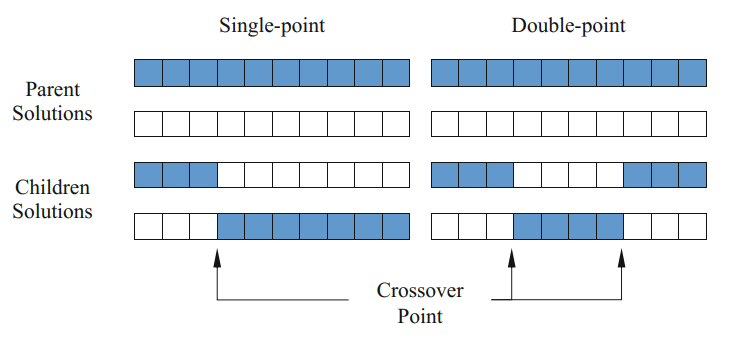}
}
\qquad
\subfigure[mutations in children solutions. 
]{
\hspace{0pt}
\includegraphics[width=.3\textwidth]{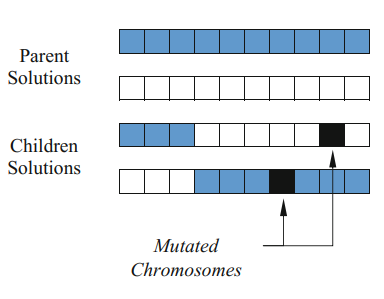}
}
\vspace*{-8pt}
\caption{Figures from \cite{genetic_alg} show (a) how selected solutions are combined for future generations; (b) mutations in children solutions. 'Chromosomes' refer to individual model parameters.}
\vspace{-10pt}
\label{fig:stoch2}
\end{figure}

The genetic algorithm is inspired from the Darwinian theory of evolution. Given a set of models with random parameters (referred to as the population), each model is tested and given a fitness score which determines its likelihood to pass its parameters on to the next generation. Those that pass have their parameters combined with others to create 'children'. In addition, children can have 'mutations' on their parameters, which is represented as a small stochastic change \cite{genetic_alg}.

Overall, the algorithm is composed of several intermediate steps: fitness scoring, parent selection, parent combination, and mutation. There are many proposed strategies for each sub-step. For example, a popular algorithm for parent selection compares individual model fitness scores to the total sum of scores to get the probability that the model is selected. In addition to these steps, other changes to the algorithm have been proposed. For example, each generation could preserve the highest scoring members of the population to live on. This process is called elitism.

The fitness function is usually considered the most important part of the genetic algorithm. Without a well designed fitness function, the algorithm will not converge to a usable solution. When using training data with expected outcomes, a common fitness function is simply the distance between the model result and the desired outcome. Another possible fitness function is percent accuracy. For example, when scoring models on an image classification task, their fitness score could be the percent of images classified correctly. While this fitness function rewards correct classifications, it does not care about the certainty of the model result. If instead the fitness function depended on the total distance between the actual and desired result, it would then reward models that most closely resemble a certain guess.

Once the fitness function has been chosen, the algorithm will then use those scores to select parents for the future generation. Generally, the probability that a model is predicted is proportional to its fitness score or overall rank in the population. A naive selection algorithm could simply select the highest scoring model every time, but this would sharply reduce diversity in the population. Diversity separates the genetic algorithm from a random search or gradient descent by exploring multiple local minima simultaneously. Therefore, it is important that even the worst performing models have a chance to be selected.

Given two or more selected parent models, the combination step of the genetic algorithm will create one or more child models for the next generation. Combination algorithms can have a predefined number of 'crossings' that subdivide the parent model parameters. Each subdivision is made up exclusively of a single parents parameters, alternating which parent provides its parameters. The exact location of the crossing is usually random. Figure 2 shows a visualization of the single and double-point crossing combination algorithms. The combination algorithm can also randomly select one of the parents for every parameter.

Mutations allow genetic algorithms to search for ideal solutions outside of the original population pool. However, the rate of mutation can have drastic effects on the efficacy of the algorithm. If the mutation rate is very high, the algorithm is similar to a random search.

\section{Methods}

All experiments were run on the University of Pittsburgh's Center for Research Computing (CRC), using the shared memory processing (SMP), message passing interface (MPI), and GPU Clusters. Unfortunately, due to the permissions given to users, specific power results could not be gathered. However, CRC gives the specifications of all hardware in each cluster, allowing for the values of thermal design power (TDP) to be obtained. Thermal design power is the power consumed at the theoretical max load of a processor \cite{ThermalDesignPower}. This assumption, while not ideal, can give us a general idea about how to power scales as the number of processors increase.

\subsection{Logistic regression with SGD}
For logistic regression, several different tests were run in order to compare various training methods and configurations. The two configurations, MP and MPI, represent the usage of either OpenMP on the SMP cluster or MPI on the MPI Cluster respectively. For training, asynchronous (A) and synchronous training were performed in both clusters. Each case was run on 1, 2, 4, 8, and 16 threads, with a serial baseline for each cluster.  The MPI Cluster used an Intel Xeon Gold 6342 Processor.

The logistic regression model with SGD training was implemented with C++, and OpenMP and OpenMPI were used for shared-memory and distributed-memory architecture respectively. We consider four different parallel methods (synchronous shared-memory, asynchronous shared-memory, synchronous distributed, asynchronous distributed) to speed up model training. The gisette dataset from LIBSVM is used as training data. The training data are feature-wisely scaled to [-1,1]. The data size is 7,000 and the feature size is 5,000. The batch size is 256. 

The shared-memory parallel methods use the OpenMP library. In the synchronous shared-memory method (MP), we speed up vector multiplication in each iteration with "\#pragma omp parallel for" and "\#pragma omp parallel for reduction(+: cp)". We do not use batch-size parallel computing to avoid race conditions in the model update. In the asynchronous shared-memory parallel methods (MPA), we use " \#pragma omp parallel for schedule(static, 1) " to parallel each update iteration. In the model update step, we use "\#pragma omp critical" to prevent race conditions and force multiple threads to update the model one by one.

In the distributed parallel methods, we use the OpenMPI library. Before model training, we first need to (virtually) distribute training datasets to each node. In the synchronous distributed method (MPI), each client samples part of the training data in the current iteration and calculates the gradients. A server then aggregates these gradients from each client and updates the model with the overall gradient. Finally, the server sends this updated model to each processing node. In the asynchronous distributed method (MPIA), we do not require synchronization communication. When the server receives a stochastic gradient, it updates the model immediately and only sends the current model back to the processing node that calculated that gradient. 

\subsection{Convolutional Neural Networks with SGD}

 The CNN models were implemented with PyTorch using sgd as an optimizer, and we added GPU acceleration with the Cuda library to speed up training. CNN models typically have a large number of parameters, which require intensive computational power and memory to train. As the size of models and datasets continue to grow, the training process can become computationally prohibitive, making it difficult or impossible to train these models on a CPU alone. It makes GPUs an ideal choice due to their parallel processing capabilities and speed in performing matrix operations. In this paper, we compare the performance of CNN model training with CPU or with GPU on CIFAR-10 in 10 epochs. CIFAR-10 dataset includes $50,000$ training images and $10,000$ testing images. $60,000$ ($32 \times 32$) color images are classified into $10$ categories (plane, car, bird, cat, deer, dog, frog, horse, ship, truck). Learning rate is set as 3e-2. The batch size is set as 32. The model architecture is listed below. We choose cross entropy as loss. The CPU used was a Intel Xeon Gold 6126 CPU from the CRC SMP Cluster, and the GPU was an NVIDIA A100 from the CRC GPU Cluster.

\begin{table} [ht]
  \centering
  \caption{ Model Architecture for CIFAR-10 \cite{huang2021efficient}}
  \label{tab:5}
\begin{tabular}{llc}
\hline Layer Type & Shape & padding\\
\hline Convolution + ReLU & $3 \times 3 \times 16 $ & 1 \\
Max Pooling & $2 \times 2$ \\
Convolution + ReLU & $3 \times 3 \times 32$ & 1\\
Max Pooling & $2 \times 2$ \\
Convolution + ReLU & $3 \times 3 \times 64$ & 1\\
Max Pooling & $2 \times 2$ \\
Fully Connected + ReLU & 512 \\
Fully Connected + ReLU & 64 \\
Fully Connected + ReLU & 10 \\
\hline
\end{tabular}
\end{table}

\subsection{Genetic Algorithm Library}

All genetic algorithm experiments were performed using the pca\_genetic library. It was written in C++ for the purposes of this project, and we used Catch2 for unit testing and libjpeg-turbo to read the JPEG images in the training dataset. It has been parallelized using OpenMP.

\begin{figure}[ht]
\centering
\includegraphics[width=.8\textwidth]{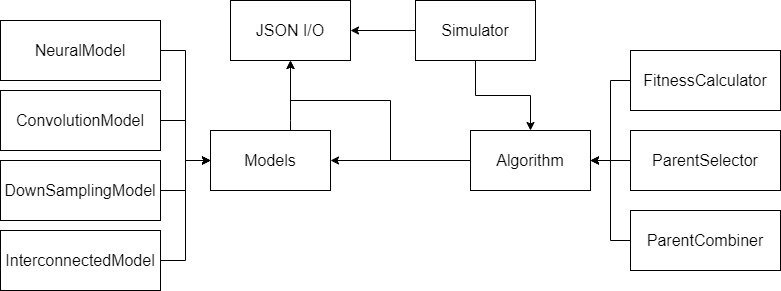}
\caption{High level architecture of the pca\_genetic algorithm. Specific model classes used in the experiments are included in the diagram}
\end{figure}

The Models package includes all of the machine learning models used for experiments along with code to interpret different types of input data. To build ANNs, the NeuralModel class was used. To build CNNs, the NeuralModel, ConvolutionModel, and DownSamplingModel were combined using the InterconnectedModel class to create the full network. The InterconnectedModel can handle an arbitrary number of models linked together in layers, so it is easily reconfigured to include multiple convolution and pooling steps. All models can represent their parameters as a vector of numbers and mutate themselves, so the genetic algorithm code can work with any model.

The genetic algorithm code include interfaces for the sub-steps of the algorithm. Fitness calculation, parent selection, and parent combination are all offloaded to individual classes. This allows the algorithm code to be easily configured, and it can swap out different algorithms for each sub-step at runtime.

The genetic algorithm will also handle several parameters that dictate the population. The mutation rate determines how about a new child model will be mutated, while the mutation size determines how large mutations can be (represented as a percentage). Generation size determines how many models exist in the population, and elitism determines what percentage of the population lives on to future generations based on their fitness rank. Finally, the offset size parameter is used to create the first generation from the model template. Each of the models in the first generation are first copied from the template, then they are mutated at a rate of 100\% and a size of the offset size.

The Simulation class contains all of the relevant information to run an experiment: an algorithm class with parameters and sub-steps, a template model class, training data, and the number of generations to run. This class will run the genetic algorithm for the provided number of generations, saving the average and best fitness of each generation in a .csv file.

A JSON interpreter was also written for the library to allow experiments to be saved as text. Every model, algorithm sub-step class, and data class can represent itself as a JSON string. This allows the Simulation class to read all of its relevant information from a JSON file before executing itself. A PyThon script was also written to generate these JSON files and execute them automatically, saving the results in .csv files. This allows experiments that test several values for an algorithm parameter to run automatically. Finally, another PyThon script was written that converts the resulting .csv files into graphs and compiles them into a PDF.

There are several possible additional features to add to pca\_genetic for future work. Currently, pca\_genetic is only running shared-memory parallelism, but it could be extended to a distributed network fairly easily. Since it can interpret JSON, processing nodes could send the bulk of messages as JSON text. For example, each processing node could be sent a copy of the training data and a portion of the population, and it can then report each models fitness to a central server node. The server node can then update the population based on the fitness data. This idea could be extended to a synchronous and asynchronous algorithm similar to SGD training. Also, some research on the genetic algorithm suggest adding additional sub-steps \cite{elitism}.

We used this library to run a few different experiments. First, we created three different models to process images from the MINST dataset of handwritten digits: A 2-layer ANN, a 3-layer ANN, and a CNN. Each image is 28x28 pixels, so each model is designed to take that many input numbers. Each model generates 10 numbers representing the confidence that the image is the corresponding digit. The highest number is considered the models guess. To find parallel efficiency, we ran a short simulation with each model and average the time to compute each generation. We reran these simulations with different numbers of processors and compared the average time to the single thread baseline. We also ran these efficiency simulations again with varying numbers of input images.

\section{Results}
This section will go into detail about the results gained from the various tests.

\begin{table}[!ht]
    \centering
    \label{tbl:CNNtrain}
    \caption{Accuracy, time. and energy of training a CNN using CPU vs. GPU on CIFAT-10}
    \begin{tabular}{|l|l|l|l|}
    \hline
        Device & Accuracy (\%) & Time (s) & Energy (J) \\ \hline
        CPU & 47 & 27978.98796 & 3497373.495 \\ \hline
        GPU & 48 & 327.3535366 & 130941.4146 \\ \hline
    \end{tabular}
\end{table}

\begin{figure}[h]
\label{fig:logreg}
\centering
\includegraphics[width=.7\textwidth]{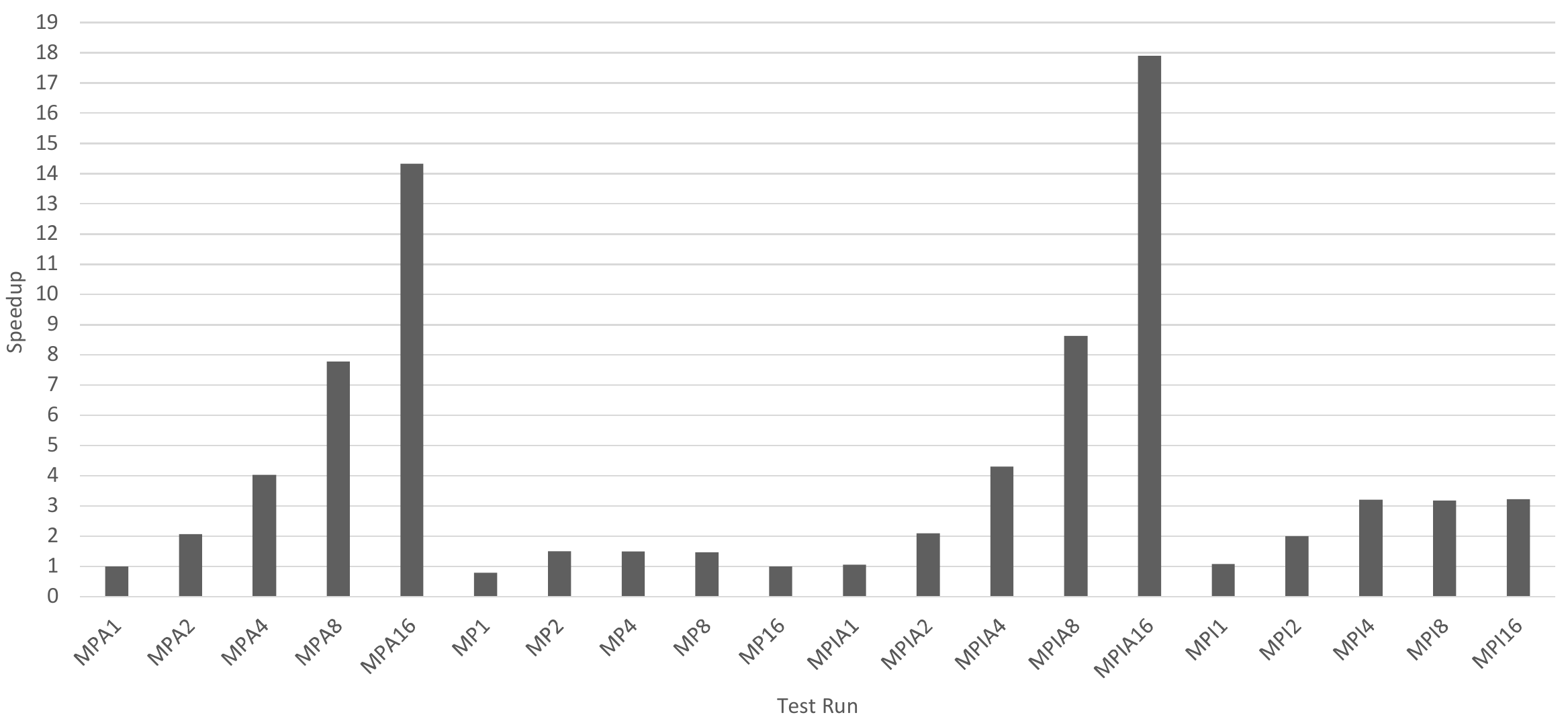}
\caption{Speedup gained using MPI and OpenMP, as well as asynchronous and synchronous training on gisette datasets}
\end{figure}

\begin{figure}[h]
\label{fig:enereg}
\centering
\includegraphics[width=.7\textwidth]{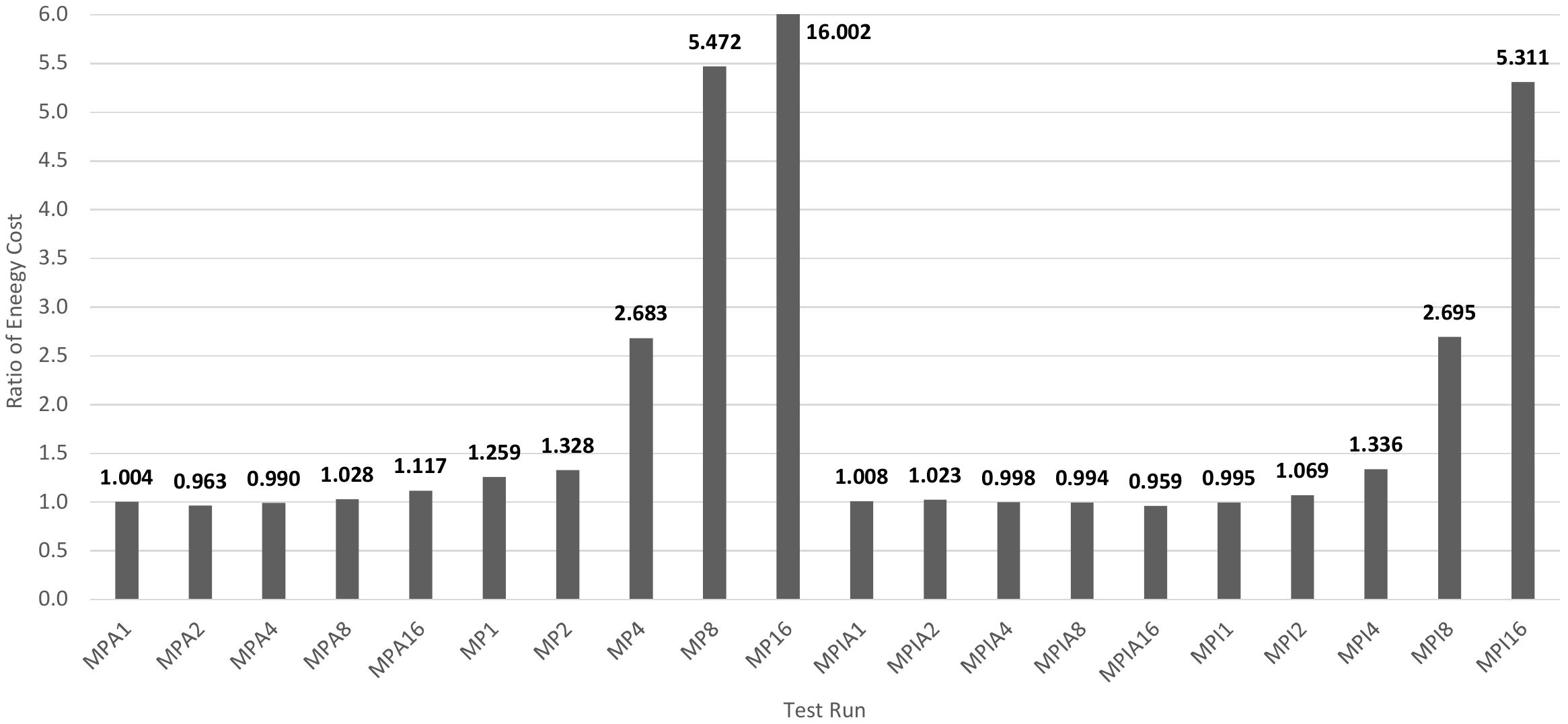}
\caption{Ratio of energy costs of MPI and OpenMP vs. serial baseline, with asynchronous and synchronous training on gisette datasets}
\end{figure}

\begin{table}[!ht]
    \centering
    \label{tbl:sync_loss}
    \caption{Final Loss results after 250 epochs for synchronous training}
    \begin{tabular}{|l|l|l|l|l|l|l|l|l|l|l|}
    \hline
        Test & MP1 & MP2 & MP4 & MP8 & MP16 & MPI1 & MPI2 & MPI4 & MPI8 & MPI16 \\ \hline
        Loss & 0.0183 & 0.0182 & 0.0183 & 0.0183 & 0.0182 & 0.0182 & 0.0185 & 0.0182 & 0.0177 & 0.0188 \\ \hline
    \end{tabular}
    
\end{table}

\begin{table}[!ht]
    \centering
    \label{tbl:async_loss}
    \caption{Final Loss results after 250 epochs for asynchronous training}
    \begin{tabular}{|l|l|l|l|l|l|l|l|l|l|l|}
    \hline
        Test & MPA1 & MPA2 & MPA4 & MPA8 & MPA16 & MPIA1 & MPIA2 & MPIA4 & MPIA8 & MPIA16 \\ \hline
        Loss & 0.0182 & 0.0194 & 0.0212 & 0.0219 & 0.0242 & 0.0182 & 0.0194 & 0.0212 & 0.0219 & 0.0242 \\ \hline
    \end{tabular}
    
\end{table}

\subsection{Logistic regression with SGD}

 The results for the speedup, as shown in Figure \ref{fig:logreg}, depict that in terms of performance, asynchronous training outperforms its synchronous counterpart in both MP and MPI cases. However, it is important to  that the final loss for the synchronous cases are roughly the same (~0.018), but it scales upwards for the asynchronous cases (~0.024 at 16 threads). These results can be seen in Tables III
 and IV.
 For some of the cases, such as MPIA4-MPIA16, speedup beyond the number of threads was obtained. This superlinear speedup can most likely be attributed to the division and caching of the data in distributed setting. Since in distributed training, each processor works with less data, it is more likely to be in cache and therefore have quicker access to it.

When looking at the plot for energy costs, the outcome is similar but not quite the same. Notice that in Figure \ref{fig:enereg}, MPIA4 has lower power efficiency than its MPA4 counterpart, despite having the higher speedup. In fact, MPIA only beats out MPA at 8 and 16 processors, due to the higher superlinear speedup it obtained at that point. The plot also shows results that correlate with Figure \ref{fig:logreg}, as the power consumed by the synchronous cases are always above the serial case due to their lackluster speedup.

Overall, since synchronous training creates an overall gradient in each iteration with current models, it is more accurate. However, asynchronous training does not involve the overhead synchronization step and has better speed-up performance. Thus, these different methods represent a trade-off between speed and accuracy. Compared with shared-memory parallel computing, distributed parallel computing has better speed-up performance.

\subsection{Convolutional Neural Networks with SGD}

 In terms of power, the GPU uses more over 3x, but ends up using over 26x less energy as seen in Table II.
 These savings in energy stem for the significant reduction in time (85x) from using the GPU's hardware, which is specifically designed to handle these types of parallel problems. The table also shows a marginal increase in accuracy, which presents the significant role of GPU in machine learning training.

\begin{figure}[ht]
\centering
\subfigure[2 Layer ANN (28*28, 10)
]{
\hspace{0pt}
\includegraphics[width=.28\textwidth]{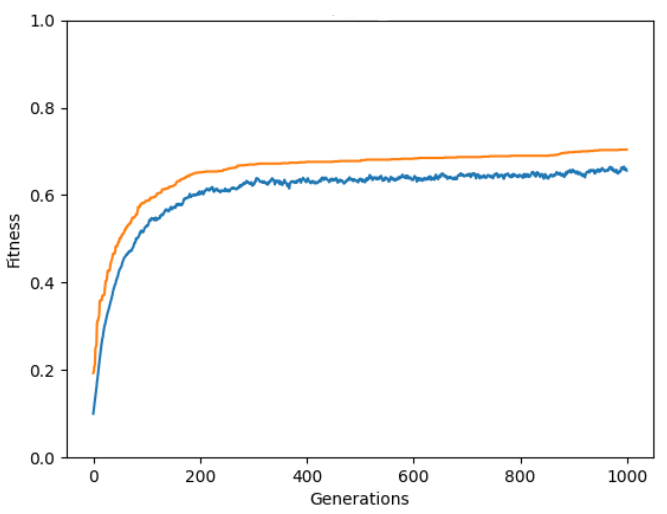}
}
\qquad
\subfigure[3 Layer ANN (28*28, 10, 10)
]{
\hspace{-20pt}
\includegraphics[width=.28\textwidth]{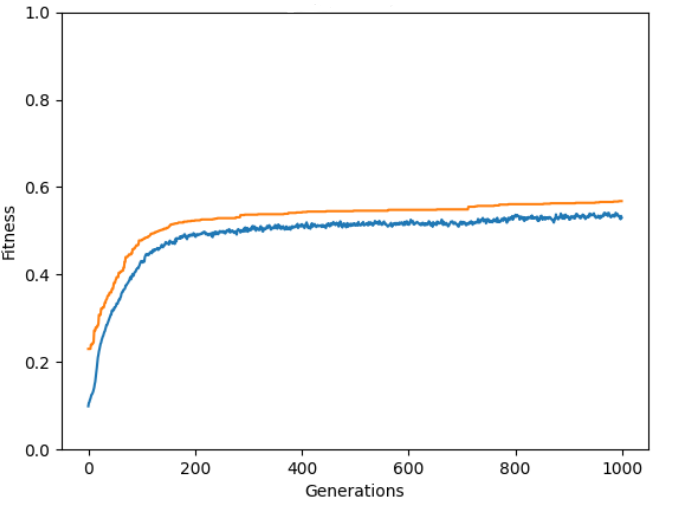}
}
\subfigure[CNN (1 feature map, 3x3 max pooling, 2 layer ANN)
]{
\hspace{0pt}
\includegraphics[width=.28\textwidth]{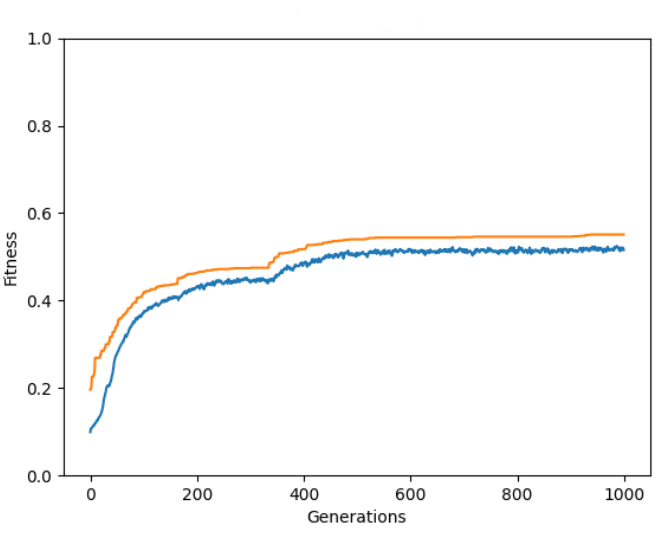}
}
\caption{Fitness values for 3 different models over 1000 generations. The 2 layer ANN reaches about 0.75 accuracy, while the other models reach about 0.6.}
\label{fig:stoch2}
\end{figure}

\begin{figure}[h]
\centering
\includegraphics[width=.5\textwidth]{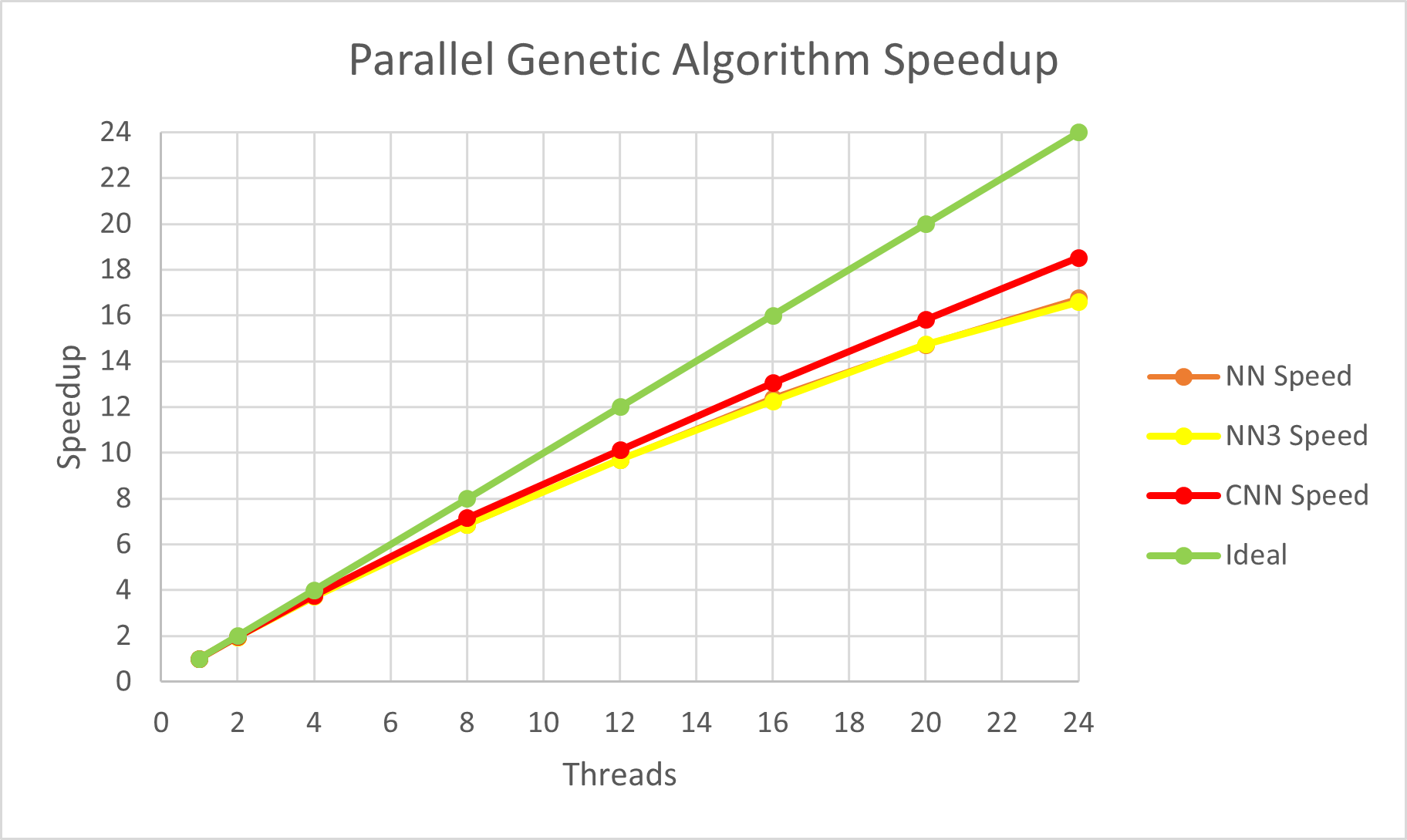}
\caption{Parallel efficiencies for the 3 different models. Each simulation was run for 100 generations with 10,000 images, and the recorded time per generation is the average of all generations.}
\label{fig:stoch2}
\end{figure}

\begin{figure}[H]
\centering
\includegraphics[width=.5\textwidth]{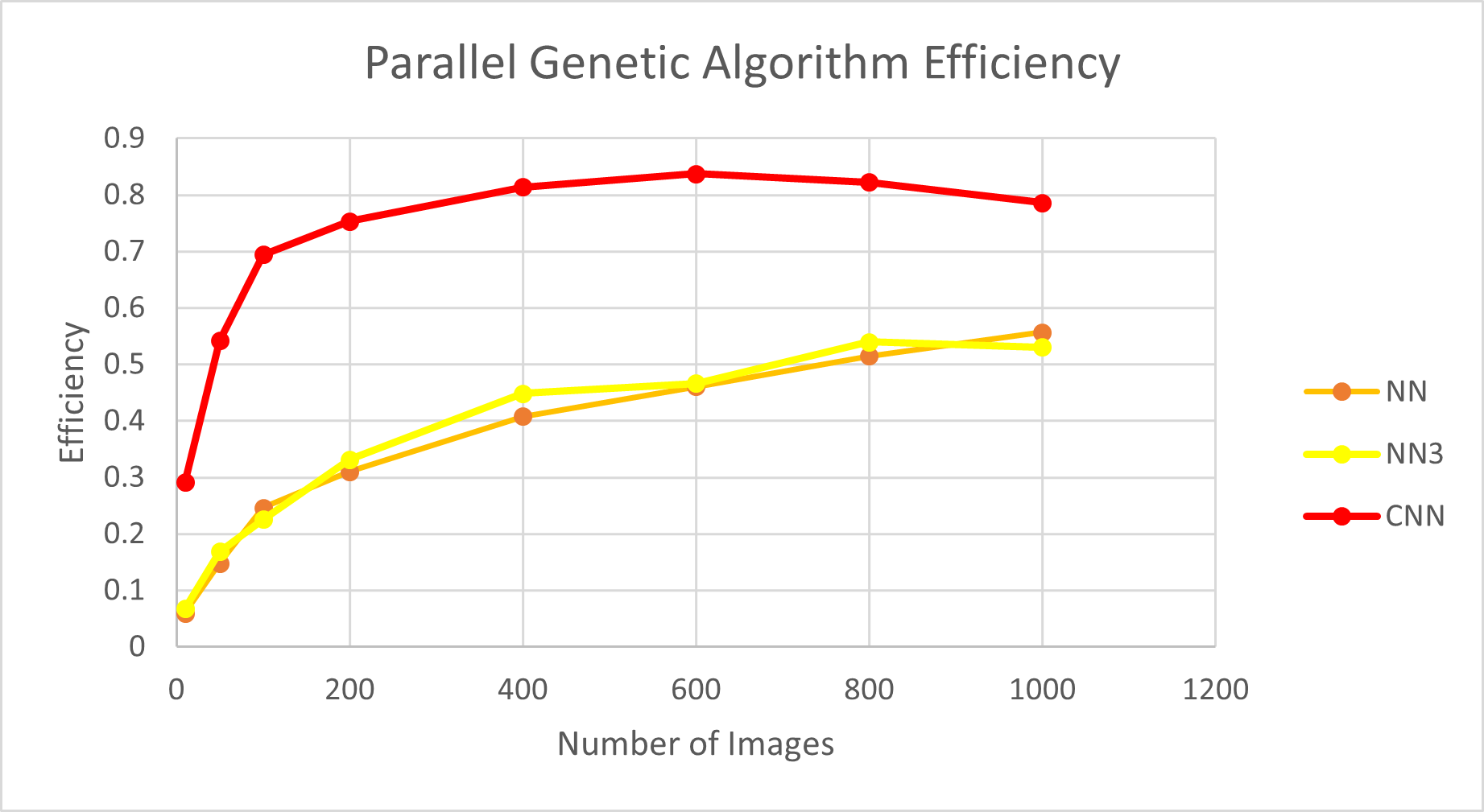}
\caption{Parallel efficiencies for the 3 different models. As the number of images increase, the parallel efficiency also rises.}
\label{fig:stoch2}
\end{figure}

\subsection{Genetic Algorithm}

When running long-term simulations, models with less complexity tend to reach a higher accuracy. This is most likely due to the smaller number of parameters overall, which reduces the search space. While the 2 layer ANN reached an accuracy of around 75\%, the 3 layer ANN and the CNN only reach about 60\%. While it is possible to reach a higher accuracy over time, it would take a very high number of generations. Each model clearly improves until around 200 generations, where further improvement is slow but not stalled. Interestingly, the CNN has a sudden sharp increase in accuracy around generation 350, but it quickly levels out again. This is most likely due to a decrease in population diversity, which forces any further improvements to depend on mutations.

Each model shows a good speedup with an increasing number of threads. As expected, the more complex CNN has a slightly higher efficiency. The additional complexity increases the granularity of each thread, which reduces the overhead of synchronizing every generation. The size of the training data set also has a large impact on efficiency. When training with fewer than 200 images, all 3 models have very low efficiencies. However, the efficiency quickly increases past this point. The CNN gains the most from the increased training data size, reaching a maximum efficiency of about 83\%.

\section{Conclusion}
In this paper, we present a parallel implementation of machine learning models (logistic regression models and CNN models) with the genetic algorithm and stochastic gradient descent (SGD) algorithm on classification tasks. We use shared-memory parallel computing, distributed parallel computing, and GPU to speed up model training. First we consider the logistic regression models with sgd as an optimizer. The experiments' results show that synchronous training is more accurate, while asynchronous training has better speed-up performance and energy consumption. In addition, compared with shared-memory parallel computing, distributed parallel computing has better speed-up and energy consumption. We then considered the CNN models with sgd as an optimizer. We trained the model on both CPU and GPU, with GPU outperforming the CPU in both power and energy. The results show that GPU reduces the training time by up to 85x while reducing the energy cost by 26x. Finally, we explored the genetic algorithm and its performance. The genetic algorithm also benefits greatly from parallelization. While the methods discussed in this paper achieved a high efficiency, further developments could be made towards distributed parallelism. An asynchronous version of the genetic algorithm could also be developed, which would make an even more efficient use of parallel resources. Overall, the algorithm shows promise in cases where SGD is not possible or effective.

\IEEEpeerreviewmaketitle
\bibliographystyle{IEEEtran}
\bibliography{reference}

\end{document}